\documentclass[aps,pre,reprint,showkeys]{revtex4-1}
\usepackage{amsmath,amssymb}
\usepackage{graphicx}
\usepackage{balance}
\usepackage{xcolor}

\newcommand{\ie}{{\it i.e.~}}
\newcommand{\eg}{{\it e.g.~}}

\newcommand{\etcn}{{\it etc\dots}}
\newcommand{\vs}{{\it vs.~}}

\newcommand{\ave}[1]{\left\langle #1 \right\rangle}

\newcommand{\elabel}[1]{\label{eq:#1}}
\newcommand{\Eref}[1]{Eq.~(\ref{eq:#1})}
\newcommand{\BEref}[1]{Equation~(\ref{eq:#1})}

\newcommand{\flabel}[1]{\label{fig:#1}}
\newcommand{\Fref}[1]{Fig.~\ref{fig:#1}}

\newcommand{\Slabel}[1]{\label{sec:#1}}
\newcommand{\Sref}[1]{Sec.~\ref{sec:#1}}

\begin{document}

\title{Spatial Point Pattern and Urban Morphology: Perspectives from Entropy,
Complexity and Networks}
\author{Hoai Nguyen \textsc{Huynh}}
\email{huynhhn@ihpc.a-star.edu.sg}
\affiliation{Institute of High Performance Computing\\Agency for Science,
Technology and Research, Singapore}

\begin{abstract}
Spatial organisation of physical form of an urban system, or city, both
manifests and influences the way its social form functions. Mathematical
quantification of the spatial pattern of a city is, therefore, important for
understanding various aspects of the system. In this work, a framework to
characterise the spatial pattern of urban locations based on the idea of entropy
maximisation is proposed. Three spatial length scales in the system with
discerning interpretations in terms of the spatial arrangement of the locations
are calculated. Using these length scales, two quantities are introduced to
quantify the system's spatial pattern, namely mass decoherence and space
decoherence, whose combination enables the comparison of different cities in the
world. The comparison reveals different types of urban morphology that could be
attributed to the cities' geographical background and development status.
\end{abstract}


\maketitle

\section{Introduction}

Cities as complex systems \cite{2005@Batty,2013@Batty,2007@Bettencourt.etal,
2010@Bettencourt.West} have been a topic of research beyond the traditional
discipline of urban studies. The idea of complexity in cities arises from the
fact that they comprise many entities interacting with one another locally and
generating global emergent patterns. Those interactions and the associated
patterns have been shown to exhibit properties
\cite[\eg][]{2007@Bettencourt.etal} similar to those observed in theoretical
models developed in the fields of Statistical Physics or Mathematics.
Quantitative tools from these fields, therefore, can be fruitfully applied
toward constructing a framework for Science of cities.

Of the many aspects of studying cities, the spatial organisation of physical
form, \ie infrastructure elements, in a city provides a fundamental
understanding of the city's way of life. Various methods have been employed to
tackle the problem of characterising spatial patterns of urban systems,
including fractal dimension \cite{1994@Batty.Longley}, land use patterns
\cite{2013@Decraene.etal,2016@Goh.etal}, street networks
\cite{2008@Barthelemy.Flammini,2014@Louf.Barthelemy}, or entropy of
population density \cite{2018@Volpati.Barthelemy}. Among them, percolation
has proved to be a powerful and useful tool to study urban morphology
\cite{1998@Makse.etal}. In recent years, percolation method has becoming
increasingly popular in analysing the spatial organisation of places in urban
systems at various scales, from city \cite{2018@Huynh.etal,2019@Huynh}, to
nation \cite{2016@Arcaute.etal} and inter-country level
\cite{2016@FLuschnik.etal}. The application of percolation in such studies has
so far been mainly concerned with studying the evolution of the giant cluster
formed when the distance threshold $\rho$ for inter-point interaction 
\footnote{a pair of points belong to the same cluster if and only if the
distance between them is not more than the threshold $\rho$, \ie one can hop
from one point in the cluster to any other points in the same cluster after
jumping a finite number of times, each time a distance no more than $\rho$.}
changes. The growth of such cluster involves a transition from a segregate state
where points are disconnected to an aggregate state in which a path exists
between a pair of points located at opposite ends of the system. The
identification of such transition regime is normally done via rate of growth of
the giant cluster as the distance threshold increases. The profiles of such
growth can be divided into $3$ parts, namely slow growth, rapid growth and
stabilisation (\Fref{percolation_cluster_growth}). At small value of distance
threshold, most clusters are localised due to limited connections with other
nearby points. As $\rho$ increases, points can have access to farther
neighbours, making small clusters merge to form larger ones. When $\rho$ is
sufficiently large, a dominant cluster emerges and rapidly grows within a narrow
range of $\rho$, known as transition regime \cite{1994@Stauffer.Aharony}. After
this regime, the dominant cluster, also called a giant cluster, starts to
stabilise as it has already grasped most of the points and only grows slowly
until no further expansion is possible, \ie all points now belong to a single,
unified cluster with a path existing between any pair of points in the set.

\begin{figure}
\centering
\includegraphics[width=\columnwidth]{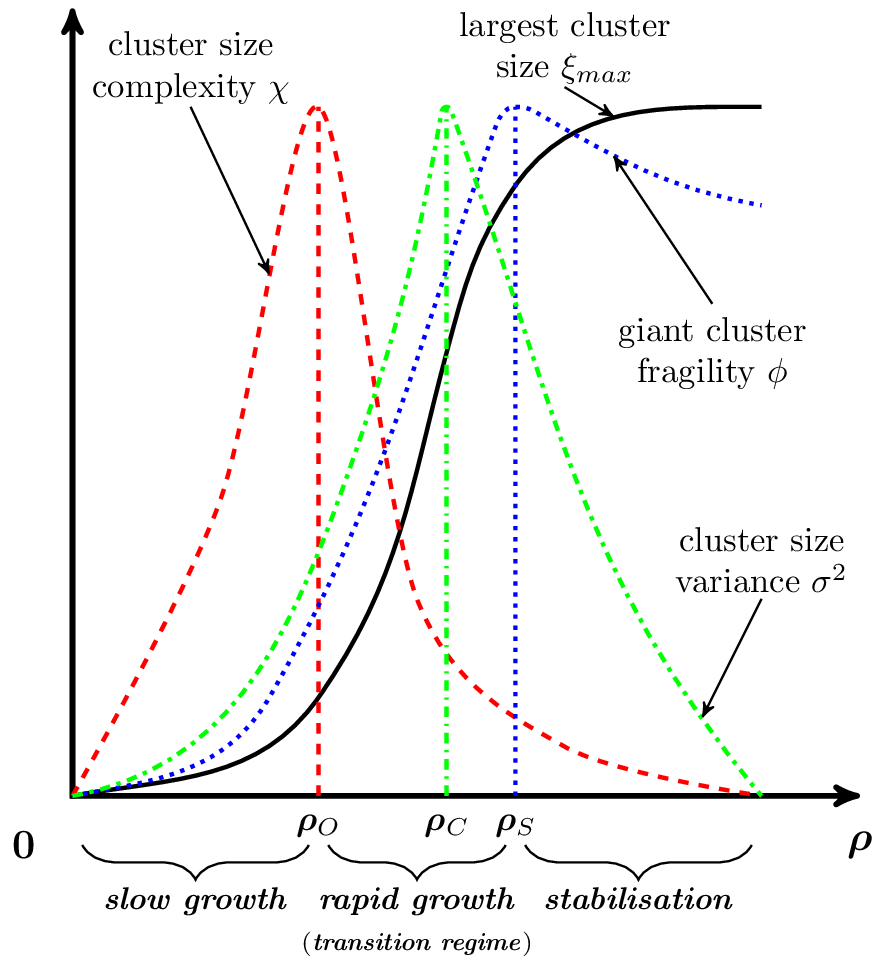}
\caption{\flabel{percolation_cluster_growth}Growth of largest cluster in a set
of points as the range of interaction among points increases. Three stages of
growth could be observed, namely slow growth, rapid growth and stabilisation. As
will be discussed in the text later, the window of rapid growth can be
determined by two length scales that signify the maximisation of entropy
measures of connected clusters ($\rho_O$ and red dashed curve) and robust
components of the giant cluster ($\rho_S$ and blue dotted curve) as range of
interaction $\rho$ changes. The rapid growth window also encompasses the
transition of the system from segregate to aggregate state, represented by the
percolation threshold $\rho_C$ at which the sizes of the clusters are most
diverse (green dash-dot curve).}
\end{figure}

Traditionally, theoretical study of percolation on regularly spaced lattices
provides procedures to quantify the transition regime and characterise it in the
framework of universality classes \cite{1994@Stauffer.Aharony,1999@Stanley}.
Extending to continuous space, continuum percolation theory relaxes the position
of points and studies their properties, including the conditions for existence
of the giant cluster under different settings
\cite{1961@Gilbert,1996@Meester.Roy}. While a number of studies have been
devoted to estimate the value of percolation threshold, especially in
thermodynamic limit for theoretical systems
\cite[\eg][]{1997@Rintoul.Torquato,2012@Mertens.Moore}, much less focus has been
put on determining the transition in a finite set of points, which could appear
very fuzzy, especially in real data. In the context of urban studies, some
measures have been applied to investigate the percolation transition in road
networks \cite{2017@Molinero.etal}, yet the transition regime in finite systems
remains largely unexplored. Such result is particularly useful for practical
applications like quantitative urban morphology, where data are always bounded.

This study, therefore, aims to present a framework to examine the transition in
the context of continuum percolation of a finite set of points, by identifying
different length scales associated with different states of the system as the
distance threshold $\rho$ changes and combining them to quantify the transition.
In the remaining of this paper, these length scales are calculated in
\Sref{length_scales}, which will be employed to characterise the transition
growth of giant cluster via two measures of mass and space decoherence. The two
measures will be used to assess different real urban systems in
\Sref{urban_application}. Finally, discussions and summary are offered in
\Sref{discussions}

\section{Length scales in continuum percolation process}
\Slabel{length_scales}

\subsection{Critical distance threshold}
Drawing upon the an important property of percolation that physical quantities
(\eg correlation length) diverge, \ie lack of characteristic size, at the
critical point, it could be paralleled that the variance of cluster size (number
of points in a cluster) maximises when the system experiences the most abrupt
change in its state. In other words, the values of cluster size are most spread
when the system transits across a ``critical'' point differentiating the
aggregate and segregate state in the system
\cite{1999@Tsang.Tsang,2000@Tsang.etal}.

To make things concrete, let us consider a set of $N$ points in a
two-dimensional domain $\mathbb{R}^2$. Given a distance threshold $\rho$, the
set is divided into $n$ clusters of size $\xi_i$, which sum up to $N$, \ie
\begin{equation}
\elabel{cluster_size}
\sum_{i=1}^{n}{\xi_i} = N \text{,}
\end{equation}
and whose variance is given by $\sigma^2=\ave{\xi^2}-\ave{\xi}^2$. The value of
distance threshold at which the variance $\sigma^2$ maximises is denoted
$\rho_C$ to mark the critical point in the transition of the system (see green
dash-dot curve in \Fref{percolation_cluster_growth}). This value is analogous
with the percolation threshold in the classic percolation theory. As with
percolation theory, the percolation threshold itself is not sufficient in
characterising the phase transition in the system. Rather, the manner of
transition is more important with many interesting properties. In what follows,
it will be shown that the window of transition could be characterised by
employing the measures of entropy. In particular, the measures of entropy can be
used to quantify the pattern of clusters formed at every value of distance
threshold and identify the length scales at which the entropy measures maximise.
As will be argued later, these length scales correspond to the change of state
of spatial agglomeration in the set of points.

\subsection{Measures of fragmentation and complexity of clustering
configuration}
\subsubsection{Measure of fragmentation}
For the clusters in \Eref{cluster_size}, the probability of choosing a random
point $a$ that belongs to a cluster $C_i$ of size $\xi_i$, also the probability
of picking the cluster $C_i$ itself, is simply given by the fraction of points
in that cluster, $p_i=p(a\in C_i)=\displaystyle\frac{\xi_i}{N}$. With this, we
can easily calculate the Shannon entropy of the particular cluster division in
\Eref{cluster_size}
\begin{equation}
\elabel{1stentropy}
S = -\sum_{i=1}^{n}{p_i\log{p_i}}
= -\sum_{i=1}^{n}{\frac{\xi_i}{N}\log{\frac{\xi_i}{N}}} \text{.}
\end{equation}

It could be seen from \Eref{1stentropy} that when there is a dominant cluster
$C_{i^\star}$ of very large size alongside several tiny clusters of vanishingly
small sizes (which are yet to be absorbed into the giant cluster), the entropy
is close to $0$ since $\displaystyle\log{\frac{\xi_{i^\star}}{N}}\approx0$ and
$\displaystyle\frac{\xi_i}{N}\approx0$, $\forall i\neq i^\star$. This reflects
the state of division that the set of $N$ points is barely fragmented, where
most of them belong to a single, unified cluster. On the other hand, it is a
well-known fact for Shannon entropy formula that given $n$ events, the
respective entropy is maximised when each of them takes place with equal
probability $\displaystyle\frac{1}{n}$, which simply yields
\begin{equation}
\elabel{entropy_equal_division}
\max{(S)} = -\sum_{i=1}^{n}{\frac{1}{n}\log{\frac{1}{n}}} = \log{n} \text{,}
\end{equation}
\ie the scenario of dividing the set of $N$ points into $n$ equal clusters. This
is the state of maximal uncertainty since any of the clusters can be picked with
equal probability. \BEref{entropy_equal_division} also indicates that the upper
bound of entropy measure for $n$ events increases with the number of events.
This points to the fact that maximal possible entropy in the system is
$S_{max}=\log{N}$ when there are $N$ clusters, each of size $1$ and being picked
with equal probability $\displaystyle\frac{1}{N}$. This corresponds to a state
of being totally fragmented when each point forms its own cluster. In other
words, the Shannon entropy in \Eref{1stentropy} can be interpreted as measure of
fragmentation in the set of $N$ points. This is considered ``first-order''
measure in the sense that the formula operates directly on the size fraction of
the individual clusters. In the following, we consider another measure that
operates on the size distribution of the clusters.

\subsubsection{Measure of complexity and onset of giant cluster formation}
We again consider a set of $N$ points being divided into $n$ clusters of size
$\xi_i$, which consist of points within distance $\rho$ of one another. Let's
denote $m(\xi)$ the number of clusters having size $\xi$ so that we have
\begin{equation}
\sum_{\xi=1}^{\xi_{max}}{m(\xi)\xi} = N \text{,}
\end{equation}
in which $\xi_{max}$ is the size of the largest cluster. The probability of
randomly choosing a point that belongs to a cluster of size $\xi$ is given by
$\displaystyle P(\xi)=\frac{m(\xi)\xi}{N}$. With this, the entropy of cluster
sizes could be calculated as
\begin{equation}
\elabel{cluster_size_entropy}
\chi = -\sum_{\xi=1}^{\xi_{max}}{P(\xi)\log{P(\xi)}}
= -\sum_{\xi=1}^{\xi_{max}}{{\frac{m(\xi)\xi}{N}}{\log{\frac{m(\xi)\xi}{N}}}}
\text{.}
\end{equation}

When there are several tiny clusters of vanishingly small sizes alongside a
dominant cluster $C_{i^\star}$ of very large size, \ie very large $\rho$, the
entropy $\chi$ is close to $0$ since
$\displaystyle\log{\frac{m(\xi_{i^\star})\xi_{i^\star}}{N}}\approx0$ (for
$m(\xi_{i^\star})=1$ and $\xi_{i^\star}\approx N$) and
$\displaystyle\frac{m(\xi)\xi}{N}\approx0$, $\forall \xi\neq\xi_{i^\star}$.
At the other extreme, when every point forms a cluster of its own, \ie very
small $\rho$, the probability of choosing a random point that belongs to a
cluster of size $\xi$ is given by a Kronecker delta $P(\xi)=\delta_{1,\xi}$, for
which the entropy $\chi$ is trivially $0$. This points to the fact that at
either extreme of cluster formation, the set of points is divided into a trivial
pattern when the size of a randomly picked cluster is not uncertain, yielding
vanishing entropy measure, \ie
\begin{equation}
\begin{aligned}
\lim_{\rho\rightarrow0}{\chi} & =-P(m(\xi)=N,\xi=1)\log{P(m(\xi)=N,\xi=1)} \\
& =0\text{,} \\
\lim_{\rho\rightarrow\infty}{\chi} & =-P(m(\xi)=1,\xi=N)\log{P(m(\xi)=1,\xi=N)}
\\
& =0\text{.}
\end{aligned}
\end{equation}
From this, it can be seen that the measure of entropy $\chi$ in
\Eref{cluster_size_entropy} exhibits a maximum value at some finite value of
$\rho$ when the clusters are formed with various sizes at which the proportion
of points in different cluster sizes are most uniform. At this juncture, it
could be pictured that each point in a cluster of size $\xi$ carries a label
$\xi$ and the division of $N$ points into different label groups transits from
trivial to non-trivial and back to trivial again, as $\rho$ changes.

While the entropy $S$ defined in \Eref{1stentropy} is interpreted as the measure
of fragmentation of the clusters, the second entropy $\chi$ defined in
\Eref{cluster_size_entropy} could be interpreted as the measure of complexity of
the clusters' pattern. The pattern is simple when most of the points carry the
same label, \ie indistinguishable, whereas a more complex pattern is produced
when many labels are needed to describe the points. This complexity measure is
useful because we can employ it to mark the onset of giant cluster formation as
the value of $\rho$ changes. At small value of $\rho$, many small clusters exist
but the number of labels is limited as the largest cluster size remains small
(a fragmented pattern can be observed in top left panel of
\Fref{complexity_fragility}). When $\rho$ increases, the labels become more
diverse when more cluster sizes come to existence with the lifting of the
largest cluster size (a mixed pattern can be observed in middle left panel of
\Fref{complexity_fragility}). However, as $\rho$ progresses further, the largest
cluster starts to grow by absorbing smaller ones, reducing the number of labels
needed, and hence, decreasing the complexity $\chi$ of the clusters' pattern (a
simple pattern with a dominating cluster can be observed in bottom left panel of
\Fref{complexity_fragility}). Once the giant cluster has been formed, it
continues to (slowly) absorb other smaller clusters, further reducing the number
of labels and decreasing the complexity $\chi$, which eventually vanishes when
only a single label is needed for all the points in a single cluster. The value
of $\rho$ at which the complexity measure attains its maximum $\chi_{max}$ is
denoted $\rho_O$ to mark the onset of giant cluster formation (see red dashed
curve in \Fref{percolation_cluster_growth}), as reasoned above.

\begin{figure*}
\centering
\includegraphics[width=0.9\textwidth]{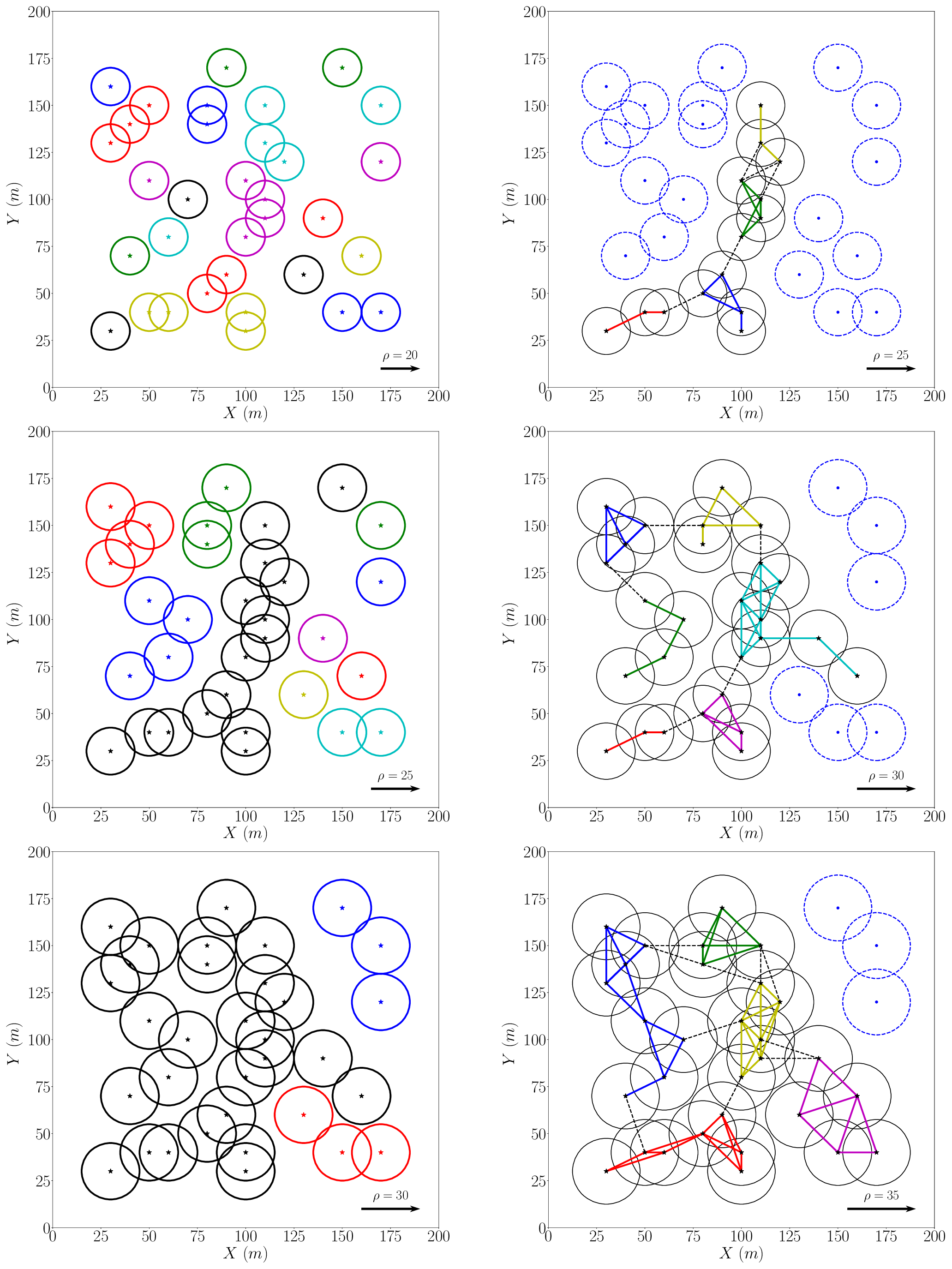}
\caption{\flabel{complexity_fragility}Illustration of complexity and fragility
measures of the percolation clusters at different values of parameter distance
(colours online). \textit{Left column:} patterns of clusters to illustrate
complexity measure. \textit{Right column:} patterns of network formed by the
giant cluster, whose communities are colour coded, to illustrate fragility
measure. \textit{Top row:} patterns slightly before the measures peak.
\textit{Middle row:} patterns when the measures peak. \textit{Bottom row:}
patterns slightly after the measures peak. For clear visuality, each points is
surrounded by a circle of radius that is equal half the value of the distance
parameter $\rho$. Any two overlapping (or touching) circles belong to the same
cluster. These circles are not to be confused with the calculation of cluster
area in \Fref{cluster_area}.}
\end{figure*}

\subsection{Measure of fragility of giant cluster}
The determination of clusters based on distance threshold indicates that there
is a path between every pair of points of a cluster. It, however, does not tell
how strongly connected the points are. In order to understand the internal
structure of a cluster, pairwise connection between every pair of points in the
cluster has to be taken into account.

To this end, a network of points' connections within a cluster could be
constructed (see right panels of \Fref{complexity_fragility}), where a link
between a pair of points, $a$ and $b$, exists if and only if their distance
$d_{ab}$ is less than the threshold $\rho$. The strength $w_{ab}$ of connection
between the pair is further taken into account in the form of the inverse of
their distance, $w_{ab}\propto d_{ab}^{-1}$, \ie a distant pair is less
connected than a closer one. With this network, it could be examined which parts
of the cluster are only weakly connected to the rest, using a community
detection method \cite{2004@Newman}, the Louvain method \cite{2008@Blondel.etal}
in particular. Once the communities within a cluster have been identified, one
can then apply the measure of fragmentation introduced in \Eref{1stentropy} to
determine the fragility of a cluster. To do this, each of the identified
communities is considered a sub-cluster within the larger cluster of interest
(whose fragility is to be quantified), and the size of the sub-cluster enters
\Eref{1stentropy} as $\xi_i$. If a cluster can be broken up into multiple
tight-knit communities, it is said to be more fragile than a cluster that
consists of only one or few closely connected communities.

Applying this to the giant cluster, it could be conceived that when the giant
cluster grows, it initially only contains a few points that are closely
connected to one another, yielding low fragility (only one or few tight-knit
communities). When the giant cluster grows further, more points are added to the
cluster, whose ties are not yet strengthened, producing multiple communities,
and hence, high fragility. This trend continues into the transition regime, with
increasing fragility. After the transition regime, most of the points are now
part of the giant cluster, slowing down the cluster's growth. At this point,
with sufficiently large value of distance threshold $\rho$, points across
different (distant) regions of the giant cluster can form links to strengthen
the ties within the community they belong to, making the cluster more robust, or
less fragile. The measure of fragmentation is useful in this case as the
distance threshold $\rho$ at which the entropy $S$ peaks, denoted $\rho_S$, is a
good indicator of the onset of stabilisation of the giant cluster (see blue
dotted curve in \Fref{percolation_cluster_growth}). This is where the giant
cluster is most fragile to be broken into components.

It should be remarked that the size of giant cluster changes with the distance
parameter. In practice, to make its fragility measure comparable across
different values of $\rho$, the sum of all the sub-clusters in \Eref{1stentropy}
is kept constant by lumping the remaining points (outside the giant cluster) as
a single cluster whose size also contributes an extra term in \Eref{1stentropy}.
This treatment also helps to take care of the scenario where multiple robust
clusters (almost fully connected, and of similar sizes) have been established
but are not yet connected to form a giant cluster. As $\rho$ increases, the
overall fragility should increase and peaks when these clusters merge, where the
newly formed giant cluster only has loosely connected components. In most other
cases, this has mild effect after the giant cluster has been formed as the
number of non-giant-cluster points is much smaller than the size of the giant
cluster itself.

As an example for illustration, right panels of \Fref{complexity_fragility} show
the networks formed by points in the giant cluster at different values of the
distance parameter $\rho$, right before, at, and after the peak value for the
fragility measure. The points in the giant cluster are surrounded with black
solid circles, whereas the others are marked with blue dashed circles. Within
the giant cluster, any pair of points whose distance is less than or equal to
distance threshold is linked by a line, which makes an edge of the corresponding
network. The communities in this network, identified using Louvain method, are
colour-coded in the plots for clarity, with intracommunity links represented as
coloured solid lines and intercommunity ones as black dashed lines. There are in
total $N=33$ points in the entire set. At $\rho=25m$, the giant cluster contains
$14$ points divided into 4 sub-clusters of size $(4,4,3,3)$, leaving $19$ points
outside. Hence, the corresponding fragility measure is $\phi_\xi=1.265$.
Similarly, at $\rho=30m$, the sub-cluster sizes are $(8,4,4,4,4,3)$, yielding
$\phi_\xi=1.895$. Finally, at $\rho=35m$, the sub-cluster sizes are
$(8,7,6,5,4)$, yielding $\phi_\xi=1.742$.

\subsection{Effective width of transition window}

The two distance scales $\rho_O$ and $\rho_S$ discussed above can be used to
determined the window of rapid growth of the giant cluster across the transition
of the system from segregate to aggregate state. Further combination with the
critical percolation distance $\rho_C$ would enable calculation of the effective
width of transition window, which characterises how the system transits from
segregate to aggregate state. To do this, it is noted that a linear growth of
the giant cluster between $\rho_O$ and $\rho_C$ should indicate a longer
effective width than that of an exponential-like growth. For this, it is useful
to use the ratio between the area $\mathcal{F}_1$ under the growth curve of
giant cluster and the change in cluster size $\xi_C-\xi_O$ as the effective
width (see \Fref{effective_windows}). Similarly, the effective width after the
critical distance, between $\rho_C$ and $\rho_S$ could be calculated in the same
manner. Subsequently, the effective width $\omega$ of the transition window
$\omega$ is simply the sum of widths both before and after the critical
percolation threshold
\begin{equation}
\elabel{effective_width}
\omega = \delta(\rho_O,\rho_C) + \delta(\rho_C,\rho_S) =
\frac{\mathcal{F}_1}{\xi_C-\xi_O} + \frac{\mathcal{F}_2}{\xi_S-\xi_C}
\text{.}
\end{equation}
This effective width is useful for it characterises the sharpness of transition
or the growth of the largest cluster, similar to the critical exponents that
characterise the divergence of a system's physical quantities (\eg correlation
length, average cluster size, \etcn) in standard percolation theory. For the
purpose of comparing different systems, a dimensionless width rescaled by the
critical percolation threshold is used
\begin{equation}
\elabel{normalised_spread}
\epsilon = \frac{\omega}{\rho_C} \text{.}
\end{equation}
It could be seen that this quantity indeed provides a measure of ``decoherence''
of relative distance among points in a set. In other words, if the points are
regularly spaced, their relative distances are mostly uniform, \ie more
coherent, yielding a small value of $\epsilon$. However, if the points are
scattered with inter-point distances ranging a wide spectrum, \ie less coherent,
the value of $\epsilon$ would surge.

\begin{figure}
\centering
\includegraphics[width=\columnwidth]{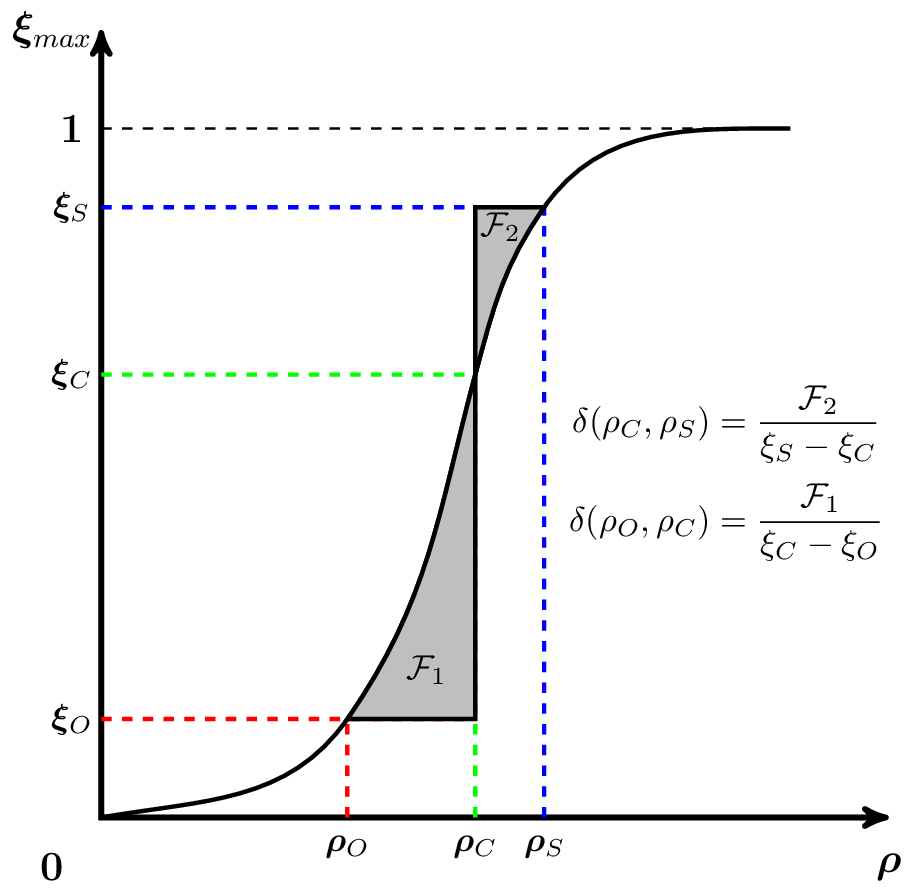}
\caption{\flabel{effective_windows}Calculation of effective width of transition
window before (marked by $\rho_O$) and after (marked by $\rho_S$) the critical
value $\rho_C$ of percolation threshold.}
\end{figure}

\subsection{Mass decoherence \vs space decoherence}
It should be noted that the discussion so far has been concerned with the
measure of size (or mass) of the clusters formed in the continuum percolation
process. As have been previously shown \cite{2018@Huynh.etal,2019@Huynh}, the
area of clusters, \ie their spatial extent, provides a different perspective to
understand the (relative) spatial arrangement of points in a domain.

The area measure $A$ of a cluster of points, formed via percolation at distance
parameter $\rho$, is defined as the union area of all the circles of radius
$\rho$ centred at those points, normalised by the area of a single such circle
(see \Fref{cluster_area}). The normalisation is needed to emphasise the fact
that the cluster area measures the compactness of the set of points, that as the
larger $\rho$ gets, the cluster area does not necessarily expand unless new
points are grasped by the cluster. This definition of cluster area is also
dimensionless and directly comparable with the size of the cluster, \ie number
of points in the cluster, which makes all the measures discussed above
conveniently extended to cluster area. It could be easily proven that the area
$A$ of a cluster (in this definition) is always smaller than its size $\xi$ due
to the non-tiling nature of circles when packed to fill space. As illustrated in
\Fref{cluster_area}, a cluster of a fixed size $\xi$ takes different values for
its area $A$ for different arrangements, with the more compact one (smaller
average nearest-neighbour distance, see lower panel of \Fref{cluster_area})
possessing smaller area.

\begin{figure}
\centering
\includegraphics[width=\columnwidth]{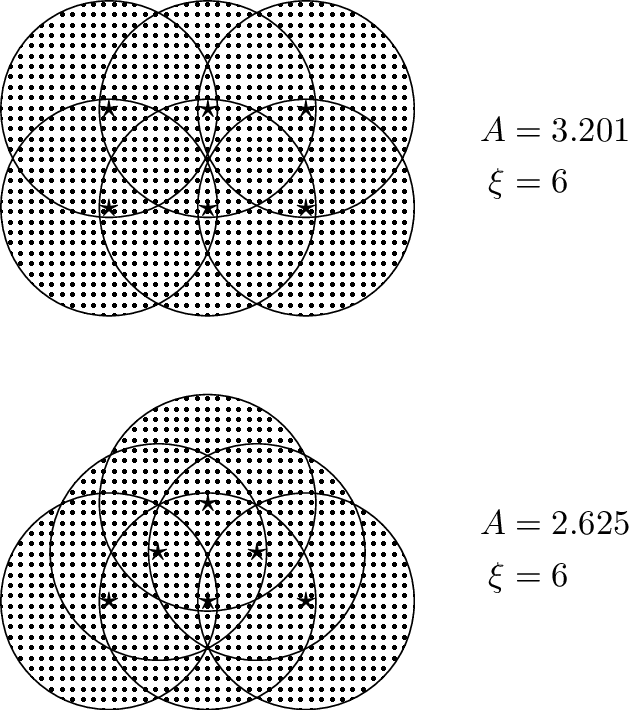}
\caption{\flabel{cluster_area}Calculation of area measure of a cluster of
points. In the upper panel, $6$ points ($\xi=6$) are arranged in a $2\times3$
lattice of spacing $d=1$, \ie the nearest-neighbour distance of every point is
$1$. In the lower one, the points' nearest-neighbour distance is either $1$ or
$\displaystyle\frac{1}{\sqrt{2}}$, in the form of an isosceles right
triangle. In both cases, the parameter distance is set at $\rho=1.1$, sufficient
to have all the points belong to a single cluster. The area measure $A$ is the
union area (dotted) of all the circles of radius $\rho$, normalised by the area
of a single circle. Simple calculations yield the values $\displaystyle A=4+
\frac{8r^2}{\pi}+\frac{6r}{\pi}\sqrt{1-r^2}-\frac{6}{\pi}\cos^{-1}{r}
\approx3.201$ and $\displaystyle A=4+\frac{4r^2}{\pi}+\frac{2r}{\pi}\sqrt{1-r^2}
+\frac{2r}{\pi}\sqrt{2-r^2}-2\cos^{-1}{r}-4\cos^{-1}{\frac{r}{\sqrt{2}}}
\approx2.625$, respectively with $\displaystyle r=\frac{d}{2\rho}=
\frac{1}{2.2}$.}
\end{figure}

The measures of size and area complement one another and their combination can
be employed to distinguish different types of spatial point distribution. In the
following discussion, $\epsilon_A$ and $\epsilon_\xi$ are used to denote the
normalised spread in \Eref{normalised_spread} calculated for cluster area and
cluster size, respectively. Hereafter, the subscripts $A$ and $\xi$ also
correspondingly denote other quantities with respect to cluster area and cluster
size. On the one hand cluster size measures the amount of points contained in
the cluster and can be interpreted as mass, on the other hand cluster area
measures the (two-dimensional) space (continuously) occupied by the points of
the cluster and can be interpreted as spatial extent. For that, we shall term
$\epsilon_A$ \emph{space decoherence} and $\epsilon_\xi$
\emph{mass decoherence}.

\section{Transition window of different types of point pattern}

To illustrate the framework developed in \Sref{length_scales}, two artificial
point patterns are analysed in details, showing the growth profiles of the
largest cluster in terms of both the cluster area and size. The effective width
of transition window is also calculated for each of the simulated point
patterns, which will then be used to calculate the decoherence measures. The
two point patterns are examples of homogenous pattern with approximately equal
nearest-neighbour distance and inhomogeneous pattern whose density of points
varies.

\subsection{Homogeneous point pattern}

As an example of homogeneous point pattern, the points are generated by randomly
displacing the sites of a regular square lattice (see top left panel of
\Fref{sim_patterns}). The amplitude of displacement applied is sufficient but
not more than the lattice spacing. The growth of both the largest cluster area
and size (middle and bottom left panels of \Fref{sim_patterns}, respectively) is
probed by gradually increasing the distance parameter $\rho$. It could be
observed that there is an abrupt increase in both the largest cluster size and
area around $\rho_C=125m$, which is the inverse of the linear density of the
point pattern, \ie square root of the number of points per unit area ($1,024$
over $4,000m\times4,000m$). The complexity measures peak rapidly when $\rho$
approaches (just before) the critical distance threshold, and drop sharply as
$\rho$ increases beyond $\rho_C$. The fragility measures of the giant cluster,
on the other hand, also peak rapidly right after $\rho_C$ but gradually decrease
further after that. If one reverses the process, tracing the fragility measures
as $\rho$ decreases, it could be seen that more (long-range) links are removed
from the network formed by the giant cluster, making it more fragile. Slightly
above the critical distance $\rho_C$, the giant cluster quickly becomes
disintegrated, broken up into smaller clusters, of which the largest cluster is
now much smaller but more robust. The window of transition for the point
pattern, marked by the peaks of complexity and fragility measures, is narrow for
both cluster size and area. The decoherence measures as defined in
\Eref{normalised_spread} are both very small with space decoherence
$\epsilon_A=0.037$ and mass decoherence $\epsilon_\xi=0.034$.

\begin{figure*}
\centering
\includegraphics[width=\textwidth]{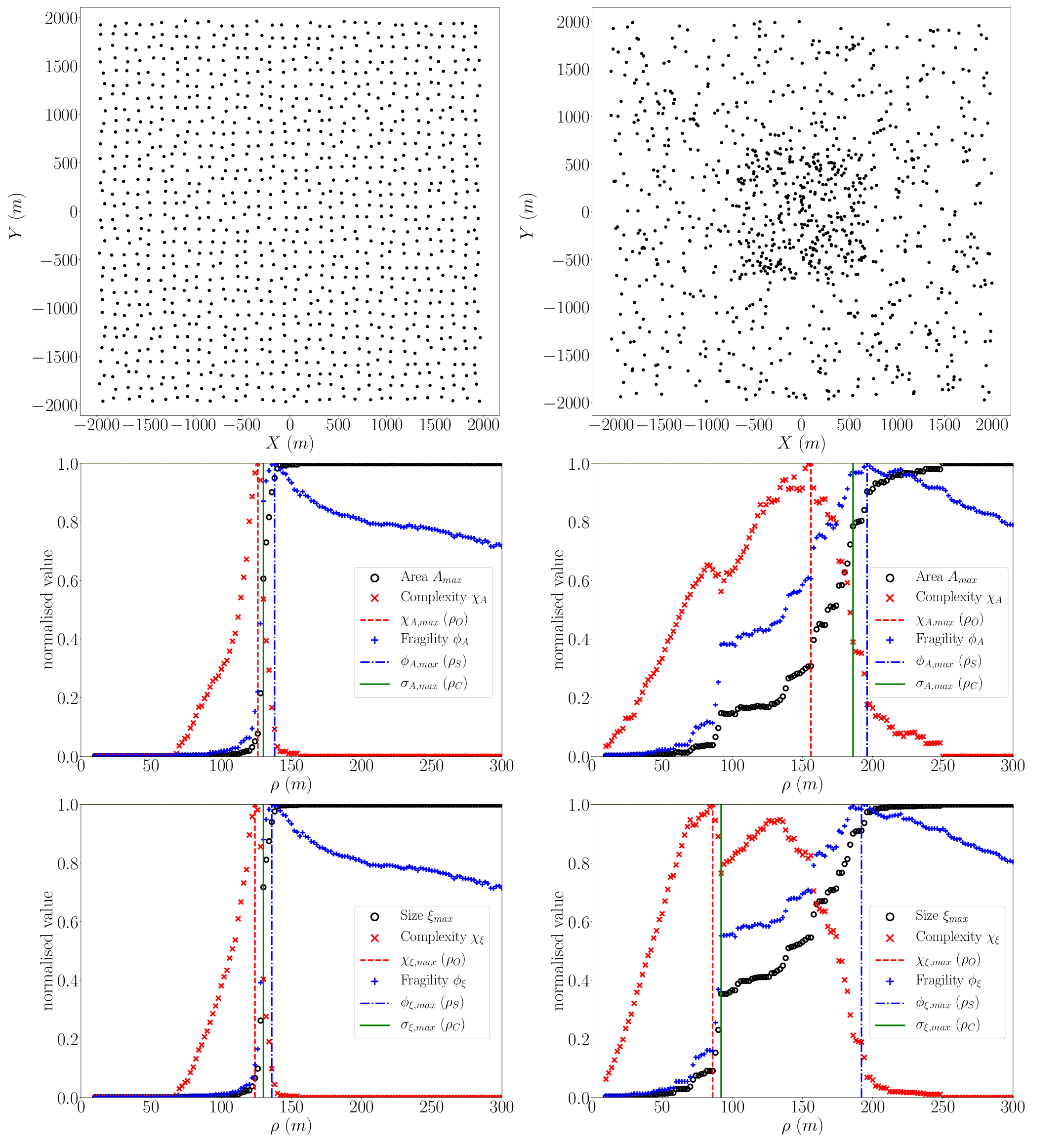}
\caption{\flabel{sim_patterns}Application of the measures in
\Sref{length_scales} to simulated point patterns (colours online). \textit{Left
column:} results for a homogeneous point pattern, which is obtained by adding
noise to a regular square lattice pattern. \textit{Right column:} results for an
inhomogeneous point pattern, whose density varies with more points concentrating
near the centre. \textit{Top row:} distribution of points in a domain of size
$4,000m\times4,000m$, with $1,024$ ($32\times32$) points on the left and $1,000$
points on the right. \textit{Middle row:} growth profile of the largest cluster
area, with vertical lines marking the length scales described in
\Sref{length_scales}. \textit{Bottom row:} growth profile of the largest cluster
size, with vertical lines marking the corresponding length scales, namely
maximum complexity (onset of giant cluster, $\rho_O$), maximum diversity
(critical distance, $\rho_C$) and maximum fragility (stabilisation of giant
cluster, $\rho_S$). The homogeneous pattern exhibits a sharp growth of both
largest cluster area and size and consequently a narrow window of transition,
whereas the inhomogeneous pattern produces a gradual growth of both largest
cluster area and size resulting in a wide window of transition.}
\end{figure*}

\subsection{Inhomogeneous point pattern}

For inhomogeneous point pattern, a simple example is obtained by having more
points concentrated in the centre region, whose density is about $4$ times more
than the rest. As the distance parameter $\rho$ increases, the growth of both
largest cluster cluster size and area shows gradual increase pattern. Since the
points tend to be clustered in the centre region (higher density, shorter
nearest-neighbour distance), the largest cluster size grows faster than the area
counterpart (see bottom and middle right panels of \Fref{sim_patterns},
respectively) as more points covering the same area compared to a lower-density
pattern. Due to the faster growth pattern, the complexity measure for cluster
size peaks earlier than that for cluster area. On the other hand, the fragility
measure of the giant cluster only peaks once many points have been (loosely)
encompassed by the cluster when $\rho$ reaches a sufficiently large value,
before declining when the giant cluster becomes more robust with more long-range
links allowed to establish at large values of $\rho$. As a result, the
transition window for the point pattern is wide for both cluster size and area,
with the latter being narrower due to the initial slower growth. The
corresponding decoherence measures are $\epsilon_A=0.114$ and
$\epsilon_\xi=0.731$, for space and mass, respectively.

\section{Application to real urban locations data}
\Slabel{urban_application}

In what follows, the two measures of space and mass decoherence are applied to a
set of $39$ cities in the world to compare the spatial patterns of their urban
morphology. The set of $39$ cities is drawn from the list of top $44$ cities
ranked by the Global Power City Index (GPCI) \cite{2018@GPCI}. The data on
spatial locations of the cities' public transport nodes were either obtained
from Open Street Map via Nextzen project \cite{Nextzen} or from General Transit
Feed Specification sources \cite{TransitFeeds}. A small number ($5$) of cities
were excluded since reliable data could not be obtained. Due to geographical
features, some cities are divided into multiple parts by large water bodies
(wide rivers or large bays or even open sea). As a result, the quantification of
spatial pattern is reported for a total of $49$ sets of points (see
\Fref{esea}).

\begin{figure}
\centering
\includegraphics[width=\columnwidth,trim=4mm 2mm 0mm 0mm]{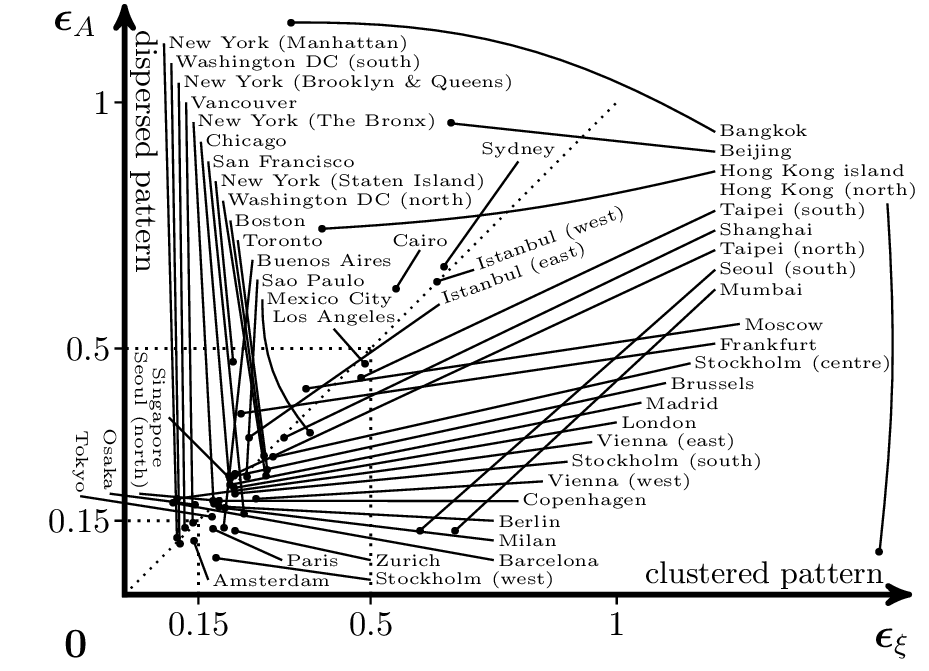}
\caption{\flabel{esea}Measures of mass and space decoherence of $49$ sets of
public transport nodes from $39$ global cities \cite{2018@GPCI}. Some cities
are split into multiple parts due to geography. General patterns could be
observed where American and European cities flock in the lower left corner of
the plot together with highly developed Asian cities from Japan, South Korea and
Singapore, while other cities are found scattered.}
\end{figure}

Using the measures of mass and space decoherence to quantify spatial patterns of
points, $3$ regions could be highlighted, namely highly coherent
($\epsilon_A,\epsilon_\xi\lesssim0.15$), coherent
($0.15\lesssim\epsilon_A,\epsilon_\xi\lesssim0.5$) and decoherent
($\epsilon_A\gtrsim0.5$ or $\epsilon_\xi\gtrsim0.5$). When the points are
decoherent, their pattern can further classified as clustered or dispersed if
one of the two measures is significantly smaller than the other (same reasoning
for $\sigma_A$ and $\sigma_\xi$ in \cite{2018@Huynh.etal}).

From the spatial pattern of $49$ sets, it could be observed that most of
American and European cities possess coherent spatial patterns with values of
$\epsilon_A$ and $\epsilon_\xi$ not exceeding $0.5$. The borderline case of Los
Angeles ($\epsilon_\xi\approx0.5$) appears to support the perception that it is
one of the most sprawling cities in the U.S. \cite{2006@Bruegmann}. Quite a
number of Asian cities also belong to this group of coherent spatial patterns,
all of which are from developed countries and ranked very high by GPCI, like
Tokyo (3), Singapore (5) or Seoul (7). The decoherent group with either
$\epsilon_A>0.5$ or $\epsilon_\xi>0.5$ contains cities mostly from developing
countries like Egypt, India or those in Southeast Asia. It is worth mentioning
that different parts of the same city divided by geography like water bodies can
possess very different morphologies. For example, the different boroughs in New
York city possess patterns ranging from high coherence of grid-like street
pattern (Manhattan) to decoherence of unplanned Staten Island. Another
interesting example is Istanbul where the Asian part east of the Bosporus strait
appears more coherent than its European portion in the west, which has been
noted in literature and could be explained by the major urban growth in
Anatolian Istanbul in the later half of last century \cite{2004@Miyamoto.etal}.
Further observations also suggest that cities known to be well-planned (and
generally ranked high by GPCI) appear to possess small decoherence values, while
the ones known for being sprawling (with tendency of lower GPCI rank) exhibit
large space and/or mass decoherence, which could either have clustered
($\epsilon_\xi\gg\epsilon_A$) or dispersed ($\epsilon_\xi\ll\epsilon_A$)
patterns.

\section{Conclusions}
\Slabel{discussions}

In summary, this work illustrates that patterns of points embedded in
two-dimensional space can be quantified using measures of complexity of the set
of points and fragility of the giant cluster in the set, formed via a continuum
percolation process. Although many sophisticated techniques have been developed
for understanding different types of spatial data \cite{2011@Cressie.Wikle}, the
general framework of percolation proves to provide a powerful toolbox to study
spatial organisation of point pattern, providing a different perspective from
that of common techniques of point pattern analysis. While point pattern
analysis mostly deals with whether a collection of points exhibit complete
spatial randomness (CSR), uniform or clustered patterns, continuum percolation
on the other hand, explores the structure of points' locations based on global
distribution, via the growth of the largest, dominating cluster in the system.
This growth is typically characterised by three stages of initial and final slow
expansion, sandwiching a rapid development region that embraces all the
interesting properties of a phase transition. Here, it is shown that window of
the transition could be determined by two length scales concerning the
complexity measure of the entire system, which marks the onset of the existence
of dominant cluster, and the component entropy of the giant cluster, which
measures its fragmentation or fragility. The former is the point at which the
complexity measure of connected clusters is maximum, while the latter is where
the giant cluster is most fragile to be broken into components, from a
perspective of network presentation of clusters. The two length scales together
with a third length scale, at which clusters are most diverse in the spirit of
critical phase transition, allow the characterisation of transition of the
system across the critical regime, in the form of decoherence measure. The
combination of mass decoherence (for amount of points accumulated) and space
decoherence (for spatial extent of points accumulated) can be employed to
quantify the pattern of a set of spatial locations, enabling comparison among
different sets. Applying this framework to the set of public transport nodes in
cities in the world from both developed and developing countries, different
types of spatial pattern can be discerned and attributed to the cities'
economical and geographical backgrounds. It is also worth mentioning that the
framework could be applied to any point data sets in urban context, \eg building
locations, road junctions \etcn, not necessarily restricted to public transport
nodes.

As a final note, the term ``complexity'' in this work is inspired by a previous
study of statistical complexity measure \cite{1989@Crutchfield.Young}, in which
two measures of metric entropy $h$ and statistical complexity $C_S$ were
calculated for non-linear dynamical systems. The measure of fragmentation $S$ in
\Eref{1stentropy} is similar to $h$, measure of randomness, which maximises at
one extreme of the system parameter and vanishes at the other; whereas the
measure of cluster size entropy $\chi$ in \Eref{cluster_size_entropy} is similar
to $C_S$, measure of complexity, which peaks at some intermediate value of the
system parameter, suggesting the idea that a system is most complex at the
interface of different states.

\bibliography{references}


\end{document}